\documentclass[superscriptaddress,twocolumn]{revtex4}
\usepackage{graphicx}
\usepackage{color}

\begin{document}

\title{Origin of a maximum of astrophysical $S$ factor 
in heavy-ion fusion reactions 
at deep subbarrier energies} 

\author{K. Hagino}
\affiliation{
Department of Physics, Tohoku University, Sendai 980-8578,  Japan}
\affiliation{Research Center for Electron Photon Science, Tohoku
University, 1-2-1 Mikamine, Sendai 982-0826, Japan}

\author{A.B. Balantekin}
\affiliation{Department of Physics, University of Wisconsin, Madison, 
WI 53706}

\author{N.W. Lwin}
\affiliation{Department of Physics, Mandalay University of Distance Education, Mandalay, 
Myanmar} 

\author{Ei Shwe Zin Thein}
\affiliation{Department of Physics, Yadanabon University, Mandalay, 
Myanmar} 


\begin{abstract}
The hindrance phenomenon of heavy-ion fusion cross sections at deep subbarrier 
energies often accompanies a maximum of an astrophysical $S$ factor 
at a threshold energy for fusion hindrance. 
We argue that this phenomenon can naturally be explained when the  
fusion excitation function is fitted with two potentials, 
with a larger (smaller) logarithmic slope at energies 
lower (higher) 
than the threshold energy. 
This analysis clearly 
suggests that the astrophysical $S$ factor provides a 
convenient tool to analyze the deep subbarrier hindrance phenomenon, 
even though the $S$ factor may have a strong energy dependence for 
heavy-ion systems unlike that for astrophysical reactions. 
\end{abstract}

\maketitle

\section{Introduction} 

Coupled-channels calculations \cite{HRK99}, 
taking into account low-lying collective 
excitations of colliding nuclei as well as several transfer channels, 
have enjoyed a great success in reproducing experimental 
fusion excitation functions for many heavy-ion systems 
at energies around the Coulomb 
barrier \cite{BT98,DHRS98,HT12,Back14}. 
The effect of channel coupling has now been well understood in terms of 
fusion barrier distributions \cite{DHRS98,RSS91,Leigh95}, that is 
fusion cross sections are given as a weighted sum of those for 
a few eigen-barriers. 

In 2002, Jiang {\it et al.} measured fusion excitation function for the 
$^{60}$Ni+$^{89}$Y system down to the 100 nb and 
discovered for the first time 
that fusion cross sections fall off much steeper 
at deep subbarrier energies 
as compared to 
a theoretical extrapolation based on the coupled-channels 
calculations \cite{Jiang02}. 
Subsequently, a similar deep subbarrier fusion hindrance has been found 
also in many other systems, see Ref. \cite{Back14} and references therein. 
Two theoretical models have been proposed in order to interpret this 
phenomenon, based either on the sudden approximation \cite{Misicu06,Simenel17} 
or on the adiabatic approximation \cite{Ichikawa07,Ichikawa15,IM13}. 
Even though the origin of the hindrance is different in these two models, 
both of them expose the importance of dynamical effects 
after two colliding nuclei touch with each other \cite{Ichikawa07-2}. 

The deep subbarrier fusion hindrance phenomenon has often been 
analyzed in terms of the astrophysical $S$ factor \cite{Back14}, 
even though the $S$ factor itself may not provide a useful tool 
for heavy-ion 
reactions -- unlike light systems in which penetration of 
the Coulomb repulsive potential makes a dominant contribution to 
reaction dynamics. 
(See also Ref. \cite{HRD03}, which was the first paper discussing 
the relation between 
the logarithmic slope of fusion excitation functions and the astrophysical 
$S$ factor). 
A somewhat surprising observation was 
that the experimental data often show a maximum in 
astrophysical $S$ factor as a function of incident 
energy \cite{Back14}. Jiang {\it et al.} argued that deep subbarrier 
hindrance sets in at the peak energy of the astrophysical $S$ 
factor \cite{Back14}. 

Even though the threshold energy so determined well follows the 
value of several global internucleus potentials at the touching 
configuration \cite{Ichikawa07-2}, the exact cause of 
the $S$ factor 
maximum 
has not yet been clarified. 
One could question if the $S$ factor could be used as a representation 
of fusion cross sections, provided that the physical meaning of the 
$S$ factor maximum is clarified. 
The aim of this paper is to address this question. 
To this end, we  
introduce 
a two-potential fit to fusion cross sections at deep subbarrier 
energies \cite{ELH12}, and show that 
the $S$ factor maximum can be naturally accounted for with this method. 
An important fact here is that the energy derivative of the astrophysical 
$S$ factor is determined by a cancellation of two terms, that is, the 
nuclear and the Coulomb contributions, and the relative importance 
between them changes precisely around the threshold energy for fusion hindrance. 

\section{Two-potential fit 
and the astrophysical $S$ factor} 

In Ref. \cite{ELH12}, we have fitted an experimental fusion excitation 
function for several systems using a single-channel 
potential model. To this end, we  
used two different Woods-Saxon 
potentials for the subbarrier and the deep subbarrier 
energy regions, which we define as the regions in which a fusion cross 
section is between 10$^{-2}$ and 10$^0$ mb, and 
below 10$^{-3}$ mb, respectively. 
Examples of the fit are shown in Figs. 1(a) and 2(a) for the 
$^{64}$Ni+$^{64}$Ni and $^{28}$Si+$^{64}$Ni systems, respectively. 
The values for the Woods-Saxon potentials are listed in Table I 
(note that we have 
used a slightly different parameter set for the 
$^{64}$Ni+$^{64}$Ni system from that shown in Ref. \cite{ELH12} 
in order to get a better fit for the astrophysical $S$ factor). 
In general, 
the surface diffuseness parameter $a$ in the Woods-Saxon potential is 
around 0.65 fm in the subbarrier region, 
however it increases to a much larger values in the 
deep subbarrier region \cite{HRD03}. 
In Ref. \cite{ELH12}, we defined the threshold energy
for the 
deep subbarrier hindrance, $E_{\rm thr}$, 
as the energy at which the fusion excitation 
functions obtained with the two potentials cross with each other. 

\begin{figure}[t]
\includegraphics[clip,width=6cm]{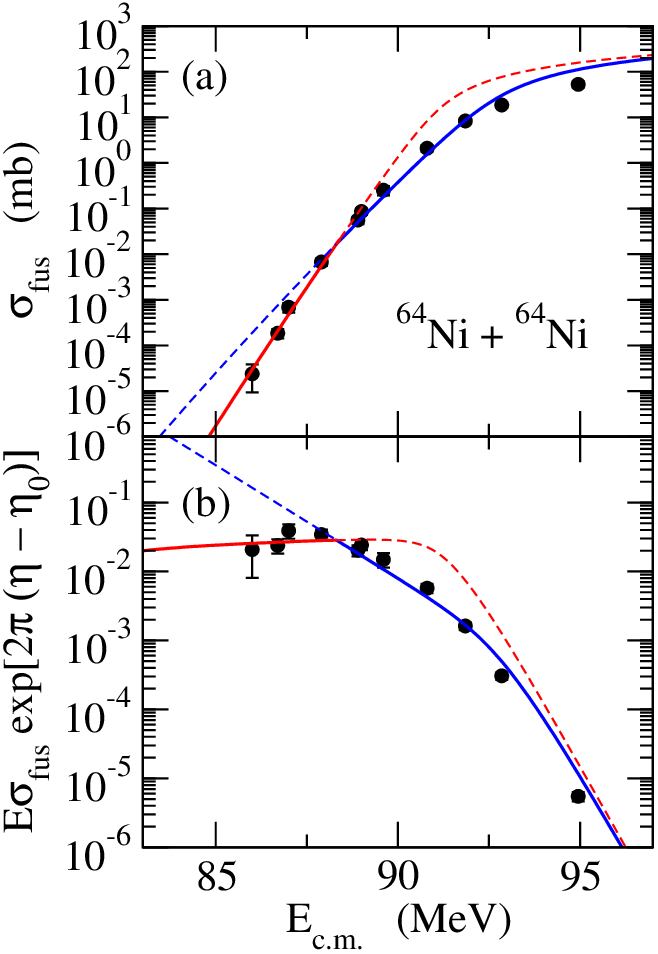}
\caption{
The fusion cross sections (the upper panel) and the astrophysical 
$S$ factor (the lower panel) for the $^{64}$Ni+$^{64}$Ni system. 
The astrophysical $S$ factor is scaled with $\eta_0$=75.23 (see text), and is given 
in units of (mb MeV). 
The solid curves denote the result of the two-potential fit, whereas the 
dashed curves show an extrapolation of the calculations 
to the region outside the fitting 
areas. The experimental data are taken from Ref. \cite{Jiang04}. 
}
\end{figure}

\begin{figure}[t]
\includegraphics[clip,width=6cm]{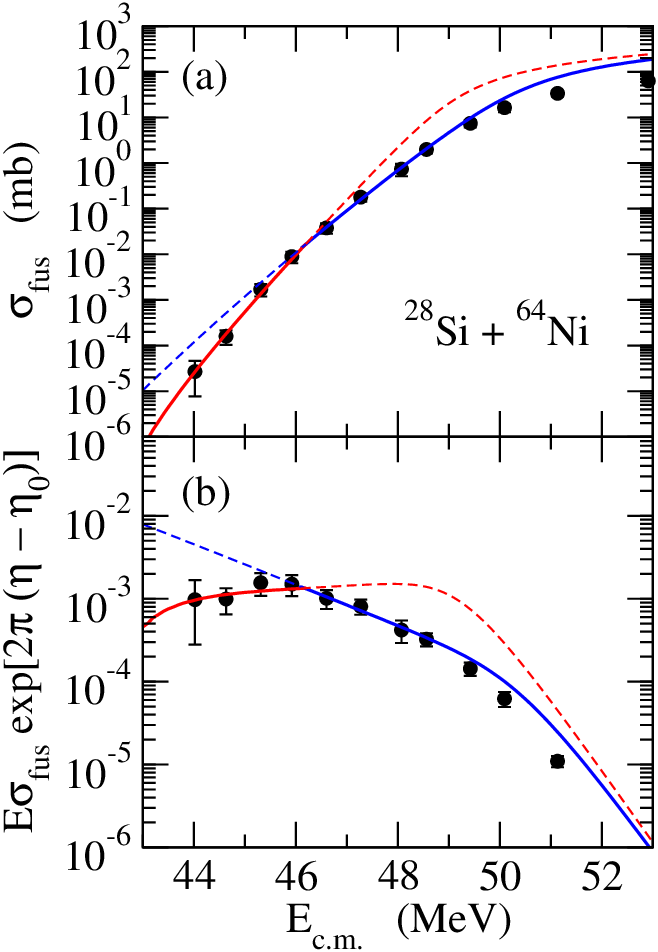}
\caption{
Same as Fig. 1, but for the $^{28}$Si+$^{64}$Ni system. 
The astrophysical $S$ factor is scaled with 
$\eta_0$=41.25. 
The experimental data are taken from Ref. \cite{Jiang06}. 
}
\end{figure}

\begin{table*}[hbt]
\caption{Parameters for the Woods-Saxon potential defined 
by $V(r)=-V_0/[1+\exp((r-R_0)/a)]$, with $R_0=r_0(A_P^{1/3}+A_T^{1/3})$,  
where $A_P$ and $A_T$ are the mass number of the projectile and the 
target nuclei, respectively. 
The subbarrier region is defined as the energy region in which 
a fusion cross section is between 10$^{-2}$ and 10$^0$ mb, while 
the deep subbarrier region is the region in which a fusion cross section 
is below 10$^{-3}$ mb. }
\begin{center}
\label{table}
\begin{tabular}{c|c|ccc}
\hline
\hline
Systems & Regions & $V_0$ (MeV) & $r_0$ (fm) & $a$ (fm) \\
\hline
$^{64}$Ni+$^{64}$Ni & subbarrier & 180 & 1.15 & 0.676 \\
                  & deep subbarrier & 98.0 & 1.1 & 1.1 \\
\hline
$^{28}$Si+$^{64}$Ni & subbarrier & 70.5 & 1.2 & 0.71 \\
                  & deep subbarrier & 46.5 & 1.19 & 0.99 \\
\hline
\hline
\end{tabular}
\end{center}
\end{table*}

The astrophysical $S$ factor, 
\begin{equation}
\tilde{S}(E)=E\sigma_{\rm fus}(E)\,e^{2\pi(\eta-\eta_0)},
\label{S}
\end{equation}
is plotted in the lower panel of Figs. 1 and 2. 
Here, $\sigma_{\rm fus}(E)$ is the fusion cross section at energy $E$, 
and $\eta=Z_PZ_Te^2/\hbar v$ is the Sommerfeld parameter, $Z_P$ and $Z_T$ 
being the atomic number of the projectile and the target, respectively, 
and $v$ being the velocity for the relative motion in the center of mass 
frame. 
For the purpose of a clear presentation, 
we scale the $S$ factor by introducing a constant $\eta_0$ in 
the exponent. 
As one can see in the figures, the energy dependence of the 
$S$ factor changes at the threshold energy, 
$E_{\rm thr}$. At energies below the threshold energy, the $S$ factor 
has a positive slope, whereas the slope becomes negative at energies 
above $E_{\rm thr}$. As a consequence, the astrophysical $S$ factor 
takes a maximum at $E=E_{\rm thr}$. 

In order to understand the energy dependence of the $S$ factor, 
let us take its first energy derivative. 
From Eq. (\ref{S}), one obtains 
\begin{equation}
\frac{1}{\tilde{S}}\,\frac{d\tilde{S}}{dE}
=L(E)-\frac{\pi \eta}{E}, 
\label{dS}
\end{equation}
where 
\begin{equation}
L(E)=\frac{1}{E\sigma_{\rm fus}}\,\frac{d}{dE}(E\sigma_{\rm fus})
=\frac{d}{dE}\ln(E\sigma_{\rm fus}), 
\end{equation}
is the logarithmic slope of a fusion excitation function \cite{Back14}. 
One can see that the energy derivative of the astrophysical $S$ factor 
consists of two terms. The first term, $L(E)$, originates from 
the nuclear potential, while the second term, $\pi\eta/E$, originates 
from the pure Coulomb interaction. 
These two terms have opposite signs, and a strong cancellation may occur. 
Figs. 3 and 4 show those contributions 
separately for the $^{64}$Ni+$^{64}$Ni and $^{28}$Si+$^{64}$Ni reactions, 
respectively. The upper and the lower panels of these figures are obtained 
with the potentials for the subbarrier and the deep subbarrier regions, 
respectively (see Table I). 
For the potentials for the subbarrier region, the logarithmic slope 
(the dashed lines) is 
relatively small, and the second term in Eq. (\ref{dS}) (the dotted lines) 
gives a larger 
contribution. The energy derivative of the $S$ factor is then negative 
at subbarrier energies. That is, the astrophysical $S$ factor is a 
decreasing function of energy in this region. 
On the other hand, for the potentials for the deep subbarrier region, 
the logarithmic slope is considerably larger than that in the subbarrier 
region, and the first term in Eq. (\ref{dS}) is comparable to the 
second term. Consequently, the energy derivative of the $S$ factor is 
slightly positive in the deep subbarrier region (see the solid lines), 
and thus the $S$ factor becomes an increasing function of energy. 
This observation is consistent with the astrophysical $S$ factors shown 
in Figs. 1 and 2. 

\begin{figure}[t]
\includegraphics[clip,width=6cm]{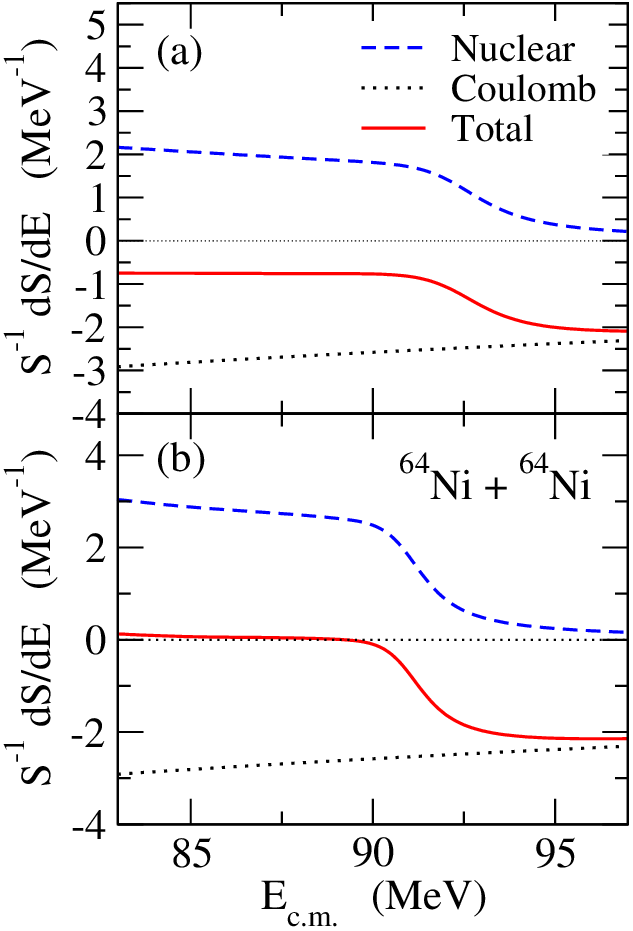}
\caption{
The first derivative of the astrophysical $S$ factor, 
$(1/\tilde{S})\,d\tilde{S}/dE$ for the 
$^{64}$Ni+$^{64}$Ni system. 
The dashed and the dotted curves show the nuclear 
and the Coulomb contributions, 
that is, the first and the second terms in Eq. (\ref{dS}), 
respectively, while the solid curves 
show the sum of these two contributions. 
The upper panel is obtained with the potential for the 
subbarrier region, while the lower panel with the potential for 
the deep subbarrier region. 
}
\end{figure}

\begin{figure}[htb]
\includegraphics[clip,width=6cm]{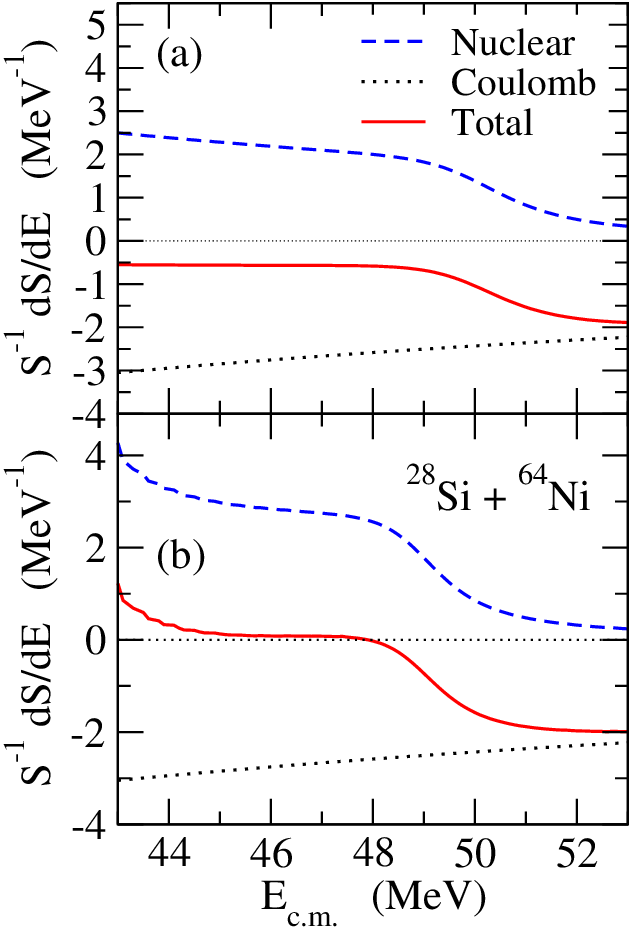}
\caption{
Same as Fig. 3, but for the $^{28}$Si+$^{64}$Ni system. 
}
\end{figure}

This analysis provides an interesting view of the astrophysical $S$ factor 
for deep subbarrier fusion reactions. As has been argued in 
Ref. \cite{Back14}, 
the maximum of astrophysical $S$ factor is well related to the deep subbarrier 
hindrance phenomenon. The hindrance of fusion cross sections leads to 
a steep falloff of fusion cross sections, and thus a large logarithmic 
slope. When the logarithmic slope becomes larger than $\pi\eta/E$, 
the $S$ factor has a positive slope as a function of energy. 
As the energy increases, 
the logarithmic slope of fusion excitation function then 
turns to a normal 
value at the threshold energy, which results in a negative slope of $S$ factor. 
This leads to a maximum in astrophysical $S$ factor at $E=E_{\rm thr}$. 

There remains the question concerning the cause of the  
change in the logarithmic 
slope of fusion excitation functions at the threshold energy, 
and the amount logarithmic slope 
changes for each system. 
In order to address the latter question, one would need microscopic 
calculations, such as those carried out in Ref. \cite{IM13} for vibrational 
excitations in a two-body system. This is beyond the scope of this paper, 
and we defer it to a future study. 
On the other hand, it is likely that 
dynamical effects after the touching configuration play an important role 
\cite{Ichikawa07-2} for the deep subbarrier fusion hindrance. 
A static effect, such as the reaction $Q$-value, has also been 
conjectured \cite{Back14}, for which the argument is that 
a fusion cross section must drop to zero at the reaction threshold 
for a system with a negative $Q$-value. 
However, this effect would be small, since the deep subbarrier fusion 
hindrance 
has been observed not only in systems with a negative $Q$ value but also 
in systems with a positive $Q$ value \cite{Back14}. 

\section{Summary} 

We discussed the relation between a maximum of astrophysical $S$ 
factor and the hindrance phenomenon in heavy-ion fusion reactions at 
deep subbarrier energies. To this end, we applied the method of 
two-potential fit to fusion cross sections. We showed that the 
logarithmic slope increases at deep subbarrier energies, which results 
in a positive energy slope in astrophysical $S$ factor, whereas the energy 
slope is negative at subbarrier energies. This leads to a maximum in 
astrophysical $S$ factor, which have been observed in many systems. 
This analysis provides a clear interpretation of the $S$ factor maximum, 
which occurs as a consequence of the change in the logarithmic slope 
of fusion excitation function at the threshold energy for the hindrance. 

The astrophysical $S$ factor has originally been introduced 
for light systems in order to remove the trivial energy dependence 
of the Coulomb penetration factor, so that an extrapolation of fusion cross 
sections down to astrophysically relevant energies can be done easily. 
Although this original purpose of introducing an astrophysical $S$ factor 
does not apply to heavy-ion 
systems, the analysis presented in this paper clearly shows that the $S$ factor 
can still be used as a convenient tool to analyze the deep subbarrier hindrance 
phenomenon, especially to identify the threshold energy for the hindrance. 

\acknowledgements

We thank the International Science Programme of Uppsala University 
for financial support for a trip of K.H. to Mandalay University 
under the project ``Research Capacity Building at Mandalay University''. 
K.H. thanks 
Profs.  Khin Swe Myint,  Than Zaw Oo, 
and Kalyar Thwe, 
and the 
Department of Physics of Mandalay Univesity for their hospitality 
during his visit. 
A.B.B. thanks the Graduate Program on Physics for the Universe (GPPU) 
of Tohoku University for financial support for his trip to Tohoku University, 
where this work was completed. The research of A.B.B. was also supported in part by the US National Science 
Foundation Grant No. PHY-1514695.

\end{document}